\documentclass{svmult}

\usepackage{makeidx}         % allows index generation
\usepackage{graphicx}        % standard LaTeX graphics tool
\usepackage{natbib}
\usepackage{multicol}        % used for the two-column index
\usepackage[bottom]{footmisc}% places footnotes at page bottom
\makeindex             % used for the subject index
\def\beq{\begin{equation}}
\def\eeq{\end{equation}}
\def\beqn{\begin{eqnarray}}
\def\eeqn{\end{eqnarray}}
\def\twiddle{\lower.9ex\rlap{$\kern-.1em\scriptstyle\sim$}}
\def\bigtwiddle{\lower1.ex\rlap{$\sim$}}
\def\gtwid{\mathrel{\raise.3ex\hbox{$>$\kern-1.05em\lower1ex\hbox{
$\sim$}}}}
\def\ltwid{\mathrel{\raise.3ex\hbox{$<$\kern-1.05em\lower1ex\hbox{
$\sim$}}}}

\begin{document}

\title*{Axion Cosmology}
\author{Pierre Sikivie}
\institute{Theoretical Physics Division, CERN, CH-1211 Gen\`eve 23, 
Switzerland\\
~~~and\\ 
Department of Physics, University of Florida, Gainesville, 
FL 32611, USA \texttt{sikivie@phys.ufl.edu}}
\maketitle

Outline:

1. Thermal axions

2. Axion field evolution

3. The domain wall problem

4. Cold axions

5. Axion miniclusters

6. Axion isocurvature perturbations

\vskip1cm

\noindent
{\it Note:}~~For background information on the Strong CP Problem, 
its resolution by an axion, and the laboratory and astrophysical 
constraints on axions, the reader is referred to the lecture notes
by Roberto Peccei and Georg Raffelt in this series.  Here, we will
be concerned only with the cosmological properties of axions.  
Different authors may use different definitions of the axion decay 
constant.  In the present notes the axion decay constant is represented 
by the symbol $f_a$ and defined by the action density for QCD plus an 
axion
\begin{eqnarray}
{\cal L}_{\rm QCD + a} &=& - {1 \over 4} G^a_{\mu\nu} G^{a\mu\nu}
+ {1 \over 2} \partial_\mu a \partial^\mu a\nonumber\\ 
&~& + \sum_q \bar{q} (i \gamma^\mu \partial_\mu - m_q) q
+ {g_s^2 \over 32 \pi^2} (\theta + {a \over f_a}) 
G^a_{\mu\nu} \tilde{G}^{a\mu\nu}
\label{QCDa}
\end{eqnarray}
where $a$ is the axion field before mixing with the $\eta$
and $\pi^0$ mesons.  Eq.~(\ref{QCDa}) uses standard notation 
for the chromomagnetic field strengths, the strong coupling 
constant and the quark fields.  The axion mass (after mixing 
with the $\eta$ and $\pi^0$ mesons) is given in terms of $f_a$ 
by 
\begin{equation}
m_a \simeq 6~\mu eV \left({10^{12}~{\rm GeV} \over f_a}\right)~~~\ .
\label{mass}
\end{equation}
$f_a$ is related to the magnitude $v_a$ of the vacuum expectation 
value that breaks the $U_{\rm PQ}(1)$ symmetry: $f_a = v_a/N$.  $N$ 
is an integer characterizing the color anomaly of $U_{\rm PQ}(1)$.  
$N=6$ in the original Peccei-Quinn-Weinberg-Wilczek axion model.  
All axion couplings are inversely proportional to $f_a$.

\section{Thermal axions}
\label{sec:1}

Axions are created and annihilated during interactions among particles in
the primordial soup.  Let us call ``thermal axions" the population of
axions established as a result of such processes, to distinguish them
from the population of ``cold axions" which we discuss later.

The number density $n_a^{\rm th}(t)$ of thermal axions solves the
Boltzmann equation \cite{KT}:
\beq
{dn_a^{\rm th} \over dt}~+~3Hn_a^{\rm th}~=
~\Gamma(n_a^{\rm eq}~-~n_a^{\rm th})
\label{Bol}
\eeq
where
\beq
\Gamma~=~\sum_i~n_i~<\sigma_i~v>
\label{rat}
\eeq
is the rate at which axions are created and annihilated,
$H(t)$ is the Hubble expansion rate, and
\beq
n_a^{\rm eq}~=~{\zeta(3) \over \pi^2}T^3
\label{equ}
\eeq
is the number density of axions at thermal equilibrium.
$\zeta(3) = 1.202.. $ is the Riemann zeta function of argument 3.
In Eq. (\ref{rat}), the sum is over processes of the type
$a~+~i \leftrightarrow 1~+~2$, where 1 and 2 are other particle
species, $n_i$ is the number density of particle species $i$, 
$\sigma_i$ is the corresponding cross-section, and the brackets 
indicate averaging over the momentum distributions of the particles 
involved.

Unless unusual events are taking place, $T \propto R^{-1}$ where 
$R(t)$ is the scale factor, and Eq. (\ref{equ}) implies therefore
\beq
{dn_a^{\rm eq} \over dt}~+~3Hn_a^{\rm eq}~= 0~~~\ .
\label{eqBol}
\eeq
Combining Eqs. (\ref{eqBol}) and (\ref{Bol}), one obtains
\beq
{d \over dt}[R^3(n_a^{\rm th} - n_a^{\rm eq})]
= - \Gamma~R^3(n_a^{\rm th} - n_a^{\rm eq})~~~\ .
\label{eqth}
\eeq
Eq. (\ref{eqth}) implies that a thermal distribution of axions is
approached exponentially fast whenever the condition
\beq
\Gamma~>~H
\label{con}
\eeq
is satisfied.  So, we have a thermal population of axions today
provided inequality (\ref{con}) prevailed for a few expansion times
at some point in the early universe, and provided the thermal
population of axions thus established did not subsequently get
diluted away by inflation, or some other cause of huge entropy
release.  

The least model-dependent processes for thermalizing axions are:
1) $a + q (\bar{q}) \leftrightarrow g + q (\bar{q})$,
2) $a + g \leftrightarrow q + \bar{q}$ and
3) $a + g \leftrightarrow g + g$.  
The corresponding diagrams are shown in Fig.~1.  These processes 
involve only the coupling of the axion to gluons, present in any
axion model [see Eq.~(\ref{QCDa})], and the coupling of quarks 
to gluons. 
\begin{figure}
\centering
\includegraphics[height=5cm]{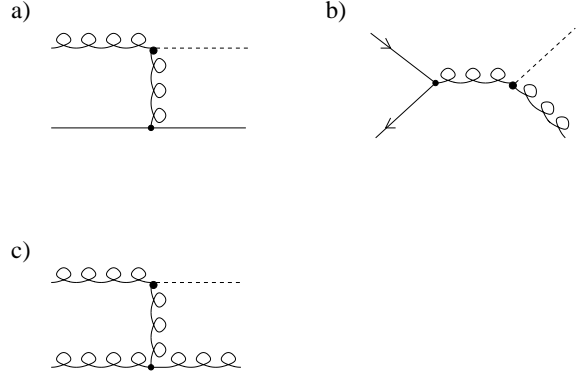}
%\picplace{5cm}{2cm} % Give the correct figure height and width in cm
%
\caption{Processes which produce thermal axions in the early universe}
\label{fig:1}       % Give a unique label
\end{figure}
A detailed treatment is given in ref. \cite{Mass02}.  We give only a
rough estimate here.  The processes of Fig.~1 have cross-sections of 
order
\beq
\sigma \sim {\alpha_s^3 \over 8 \pi^2} {1 \over f_a^2}~~~\ ,
\label{cro}
\eeq
where $\alpha_s = {g_s^2 \over 4\pi}$.  At temperatures $T>$ 1 TeV, 
the densities of quarks, antiquarks and gluons are given by
\beqn
n_q~=~n_{\bar q}~&=&~27~{\zeta(3) \over \pi^2}~T^3\nonumber\\
n_g~&=&~16~{\zeta(3) \over \pi^2}~T^3~~~\ .
\label{qde}
\eeqn
The Hubble rate is given by the Friedmann equation
\beq
H^2~=~{8 \pi G \over 3}~\rho~
=~{8 \pi G \over 3}~({\cal N}_b(T) + {7 \over 8} {\cal N}_f(T))
~{\pi^2 \over 30}~T^4~~,
\label{hub}
\eeq
where ${\cal N}_b(T)~({\cal N}_f(T))$ is the total effective
number of bosonic (fermionic) spin degrees of freedom at
temperature $T$.  For $T>$ 1 TeV, 
\beq
{\cal N} \equiv {\cal N}_b~+~{7 \over 8}~{\cal N}_f~=~107.75
\label{eff}
\eeq
if we assume no new degrees of freedom other than those of the
Standard Model plus an axion.  Combining everything and setting
$\alpha_c \simeq 0.03$, one finds
\beq
\sum_{i=1}^3~n_i~<\sigma_i v>~H^{-1}~\sim~
2~\left({10^{12}~{\rm GeV} \over f_a}\right)^2~
{T \over 10^{12}~{\rm GeV}}~~~\ .
\label{fin}
\eeq
Thus we find that the processes of Fig.~1 keep axions in 
thermal equilibrium with the primordial soup till the
temperature
\beq
T_{Ds}~\sim~5\cdot10^{11}~{\rm GeV}~
\left({f_a \over 10^{12}~{\rm GeV}}\right)^2~~~\ .
\label{tem}
\eeq
Note that the calculation is not valid when $T \gtwid v_a~=~N~f_a$,
since the PQ symmetry is restored then.  In particular, in view of
Eq. (\ref{tem}), the processes under consideration only produce a
population of thermal axions if $f_a \ltwid 2~N~10^{12}$ GeV.  

We will see in Section 4 that $f_a$ has to be less than approximately
$10^{12}$ GeV to avoid overclosing the universe with cold axions.
That bound suggests that the processes of Fig.~1 do produce a 
population of thermal axions.  We should keep in mind however 
that this thermal axion population may be wiped out by a period 
of inflation with reheat temperature less than $T_{Ds}$.  So it is
interesting to search for processes that may re-establish a thermal 
axion population later on.  We briefly discuss two such processes.

First we consider the Compton-like scattering process
$Q(\bar{Q})~+~g~\leftrightarrow ~Q(\bar{Q})~+a$.
The quark $Q$ may be a known quark or a new heavy quark.  
There are related processes in which the gluon is replaced 
by a photon or Z-boson, and/or the quark is replaced by a 
lepton.  However $Q(\bar{Q})~+~g~\leftrightarrow ~Q(\bar{Q})~+a$ 
is in some sense the least model dependent among the Compton-like
processes since, in every axion model, there is at least one colored
fermion $Q$ which carries PQ charge, and hence to which the axion
couples.  The cross-section is
\beqn
\sigma_Q~&\sim&~\alpha_s\left({m_Q \over v_a}\right)^2 {1 \over T^2}
~~~~~{\rm for}~~T>m_Q\nonumber\\
&\sim&~\alpha_s~{1 \over v_a^2}~~~~~~~~~~~~~~~~~{\rm for}~~T<m_Q~~~\ .
\label{Csc}
\eeqn
The relevant regime is when $T>m_Q$ since the $Q$ number density
is Boltzmann suppressed for $T<m_Q$.  Using $\alpha_c~=~0.05~,
~{\cal N}~=~107.75$, we have
\beq
n_Q~<\sigma_Q v>~H^{-1}~\sim
{m_Q^2 \over T~(2 \cdot 10^7~{\rm GeV})}
\left({10^{12}~{\rm GeV} \over v_a}\right)^2
\label{Cra}
\eeq
for $T>m_Q$.  So this process produces a population of thermal axions
provided:
\beq
m_Q~\gtwid~2 \cdot 10^7 {\rm GeV}~
\left({v_a \over 10^{12}~{\rm Gev}}\right)^2~~~\ .
\label{CmQ}
\eeq
The axions decouple then at a temperature $T_{DC} \sim m_Q$.

Let us also consider the process $\pi~+~\pi~\leftrightarrow~\pi~+~a$.
Since the axion necessarily mixes with the $\pi^0$, this process is 
model-independent as well.  It occurs at $T \sim 200$ MeV, after the 
QCD phase-transition but before the pions have annihilated.  The 
cross-section is of order
\beq
\sigma_\pi~\sim~{1 \over f_a^2}~~~\ .
\label{pis}
\eeq
Using ${\cal N} = 17.25$, we find
\beq
n_\pi <\sigma_\pi v> H^{-1} \sim
\left({3 \cdot 10^8~{\rm GeV} \over f_a} \right)^2
\label{pir}
\eeq
at $T \sim m_\pi$.  The $\pi~+~\pi~\leftrightarrow~\pi~+~a$ process
has the advantage of occuring very late, so that any thermal axion
population it establishes cannot be wiped out by inflation. (Inflation
occuring that late would also wipe out the baryons.)  However, 
Eq. (\ref{pir}) indicates that it is ineffective unless the 
bound $f_a > 10^9$ GeV from SN1987a is saturated.

We have seen that, under a broad set of circumstances, a population
of relic thermal axions is produced.  For $f_a > 10^9$ GeV, the axion
lifetime exceeds by many orders of magnitude the age of the universe.
Between their last decoupling, at temperature $T_D$, and today the 
thermal axion population is merely diluted and redshifted by the 
expansion of the universe.  Their present number density is
\beq
n_a^{th} (t_0) = {\zeta (3)\over \pi^2} T_D^3 
\left({R_D\over R_0}\right)^3
\label{num}
\eeq
where ${R_D\over R_0}$ is the ratio of scale factors between the time
$t_D$ of last decoupling and today.  Their average momentum is:
\beq
\langle p_a^{th} (t_0)\rangle = 
{\pi^4\over 30\zeta (3)} T_D {R_D \over R_0}
= 2.701~T_D {R_D\over R_0}\ .
\label{7.1.8}
\eeq
If $\langle p_a^{th} (t_0)\rangle >> m_a$, the energy distribution
is thermal with temperature
\beq
T_{a 0} = T_D {R_D \over R_0}\ .
\label{ate}
\eeq
If there is no inflation, nor any other form of entropy release,
from $t_D$ till the present, $T_{a 0}$ is related to the present
cosmic microwave background temperature $T_{\gamma 0} = 2.735$ K
by the conservation of entropy.  Taking account of the fact that
electron-positron annihilation occurs after neutrino decoupling,
one finds
\beq
T_{a 0} = \left({10.75 \over {\cal N}_D} \cdot
{4 \over 11}\right)^{1 \over 3} T_{\gamma 0} =
0.905 {\rm K} \left({106.75 \over {\cal N}_D}\right)^{1 \over 3}
~~~\ .
\label{axtemp}
\eeq
The average momentum of relic thermal axions is
\beq
\langle p_a^{\rm th} (t_0) \rangle = 2.1~10^{-4}{\rm eV}
\left({106.75 \over {\cal N}_D}\right)^{1 \over 3}~~~,
\eeq
and their number density is
\beq
n_a^{\rm th} (t_0) = {7.5 \over {\rm cm}^3}
\left({106.75 \over {\cal N}_D}\right)~~~\ .
\eeq

\section{Axion field evolution}

The thermal axions discussed in the previous Section are quantum
fluctuations about the average background value of the axion field.
The evolution of the average axion field, from the moment U$_{\rm PQ}(1)$
gets spontaneously broken during the PQ phase transition to the moment
the axion acquires mass during the QCD phase transition, is the topic
of this Section.

The U$_{\rm PQ}(1)$ symmetry gets spontaneously broken at a critical
temperature $T_{\rm PQ} \sim v_a$, where $v_a$ is the vacuum expectation 
value of a complex field $\phi(x)$.  The action density for this 
field is of the form
\begin{equation}
{\cal L}_\phi = {1 \over 2} \partial_\mu \phi^\dagger \partial^\mu \phi
- {\lambda \over 4}(\phi^\dagger \phi - v_a^2)^2 + ....
\label{lphi}
\end{equation}
where the dots represent interactions with other fields in the theory.
At $T> T_{\rm PQ}$, the free energy has its minimum at $\phi = 0$. At 
$T< T_{\rm PQ}$, the minimum is a circle, whose radius quickly approaches
$v_a$ as $T$ decreases.  Afterwards
\beq
\langle \phi (x)\rangle = v_a~e^{ia(x)/v_a}
\label{vev}
\eeq
where $a(x)$ is the axion field before mixing with the $\pi^0$ and $\eta$
mesons.  $a(x)$ has random initial conditions.  In particular, at two
points outside each other's causal horizon the values of $a$ are
completely uncorrelated.

It is well-known that the size of the causal horizon is hugely
modified during cosmological inflation.  Without inflation, the
size of the causal horizon is of order the age $t$ of the universe.
But, during an inflationary epoch, the causal horizon grows
exponentially fast and becomes enormous compared to $t$.  There
are two cases to consider.  Case 1: inflation occurs with reheat
temperature smaller than $T_{\rm PQ}$, and the axion field is
homogenized over enormous distances.  The subsequent evolution of
this zero momentum mode is relatively simple.  Case 2: inflation
occurs with reheat temperature larger than $T_{\rm PQ}$.  In case 2, 
in addition to the zero mode, the axion field has non zero modes, 
and carries strings and domain walls as topological defects.

The early universe is assumed to be homogeneous and isotropic.  Its
curvature is negligible.  The space-time metric can therefore be
written in the Robertson-Walker form:
\beq
-ds^2 = dt^2 - R(t)^2 d\vec{x} \cdot d\vec{x}
\label{met}
\eeq
where $\vec{x}$ are co-moving spatial coordinates and $R(t)$ is the
scale factor.  The equation of motion for $a(x)$ in this space-time is:
\beq
D_\mu \partial^\mu a(x) + V_a^\prime(a(x))
= (\partial_t^2 + 3 {\dot R\over R} \partial_t -
{1\over R^2}\nabla_x^2) a(x) + V_a^\prime(a(x)) = 0\ ,
\label{feq}
\eeq
where $V_a$ is the effective potential for the axion field, and
prime indicates a derivative with respect to $a$.  $V_a$ results from 
non-perturbative QCD effects associated with instantons \cite{tH}.
They break U$_{\rm PQ}(1)$ symmetry down to a $Z(N)$ discrete 
subgroup \cite{Sik82}.  $V_a(a)$ is therefore periodic with 
period $\Delta a = 2 \pi v_a/N = 2 \pi f_a$.  We may write 
such a potential qualitatively as 
\beq
V_a = f_a^2 m_a^2(t) [1 - \cos ({a \over f_a})]~~~\ ,
\label{Va}
\eeq
where the axion mass $m_a(t) = m_a(T(t))$ is a function of 
temperature and hence of time.  Eq.~(\ref{vev}) implies that 
the axion field has range $a \in [0, 2 \pi v_a]$.  Hence there 
are $N$ degenerate vacua.  The discrete degeneracy implies
the existence of domain walls, which will be discussed in 
Section 3.

Substituting Eq.~(\ref{Va}) into Eq.~(\ref{feq}), the equation 
of motion becomes
\beq
(\partial_t^2 + 3 {\dot R\over R} \partial_t -
{1\over R^2}\nabla_x^2) a(x) + 
m_a^2(t) f_a \sin ({a(x) \over f_a}) = 0~~~\ .
\label{eom}
\eeq
The non-perturbative QCD effects associated with instantons
have amplitudes proportional to 
\beq
e^{-{2\pi\over\alpha_c (T)}} \simeq
({\Lambda_{\rm QCD}\over T})^{11 - {2\over 3} N_f}
\label{npt}
\eeq
where $N_f$ is the number of quark flavors with mass less than $T$.
Eq. (\ref{npt}) implies that the axion mass is strongly suppressed
at temperatures large compare to the QCD scale, but turns on rather
abruptly when the temperature approaches $\Lambda_{\rm QCD}$

Because the first three terms in Eq. (\ref{eom}) are proportional
to $t^{-2}$, the axion mass is unimportant in the evolution of the
axion field until $m_a(t)$ becomes of order ${1 \over t}$.  Let us
define a critical time $t_1$:
\beq
m_a(t_1) t_1 = 1~~~\ .
\label{t_1}
\eeq
The axion mass effectively turns on at $t_1$.  $m_a(T)$ was obtained
\cite{Pres83,Abb83,Dine83} from a calculation of the effects of QCD
instantons at high temperature \cite{Gross81}.  The result is:
\beq
m_a(T) \simeq 4 \cdot 10^{-9} {\rm eV} \left(\frac{10^{12}\rm
GeV}{f_a}\right)\left(\frac{\rm GeV}{T}\right)^4
\label{maT}
\eeq
when $T$ is near $1$ GeV. The relation between $T$ and $t$ follows
from Eq.~(\ref{hub}) and $H = {1 \over 2t}$.  The total effective 
number ${\cal N}$ of thermal spin degrees of freedom is changing 
near 1 GeV temperature from a value near 60, valid above the 
quark-hadron phase transition, to a value of order 30 below 
that transition.  Using ${\cal N} \simeq 60$, one has
\beq
m_a(t) \simeq 0.7~10^{20}~\frac{1}{\rm sec}\left(\frac{t}{\rm
sec}\right)^2 \left(\frac{10^{12}\rm GeV}{f_a}\right)  \ ,
\label{mat}
\eeq
which implies:
\beq
t_1 \simeq 2\cdot 10^{-7} \mbox{sec}
\left({f_a\over 10^{12} \mbox{GeV}}\right)^{1/3} \ .
\label{t1}
\eeq
The corresponding temperature is:
\beq
T_1 \simeq 1~\mbox{GeV}
\left({10^{12} \mbox{GeV}\over f_a}\right)^{1/6}.
\label{T1}
\eeq
Eq. (\ref{mat}) implies ${d \over dt}\ln(m_a(t)) < m_a(t)$ after $t_1$.
So, at least for a short while below 1 GeV, as long as Eq. (\ref{maT})
remains valid, the axion mass changes adiabatically.  The number of 
axions is the adiabatic invariant.  Conservation of the number of 
axions after $t_1$ allows us to estimate the energy density of axions
today from an estimate of their number density at $t_1$.  When the
temperature drops well below 1 GeV, the dilute instanton gas calculations
which yield Eq. (\ref{maT}) are no longer reliable.  Complicated things
happen, such as the confinement and chiral symmetry breaking phase 
transitions.  However, because $m_a >> H$ then, it is reasonable to 
expect the number of axions to be conserved, at least in order of 
magnitude.

\subsection{Zero mode}

In case 1, where inflation occurs after the PQ phase transition, the
axion field is homogenized over enormous distances.  Eq. (\ref{eom})
becomes \cite{Pres83,Abb83,Dine83}
\beq
\left({d^2 \over dt^2} + {3 \over 2t} {d \over dt}\right) a(t)
+ m_a^2(t) f_a \sin \left({a(t) \over f_a} \right)= 0~~,
\label{zeq}
\eeq
where we used $R(t) \propto \sqrt{t}$.  For $t << t_1$, we may
neglect $m_a$.  The solution is then
\beq
a(t) = a_0 + a_{1 \over 2} t^{-{1 \over 2}}
\label{ear}
\eeq
where $a_0$ and $a_{1 \over 2}$ are constants.  Eq. (\ref{ear}) implies
that the expansion of the universe slows the axion field down to a
constant value.

When $t$ approaches $t_1$, the axion field starts oscillating in response
to the turn on of the axion mass.  We will assume that the initial value
of $a$ is sufficiently small that $f_a \sin(a/f_a) \simeq a$. Let's define
$\psi$:
\beq
a(t) \equiv t^{-{3 \over 4}} \psi(t)~~~\ .
\label{psi}
\eeq
The equation for $\psi(t)$ is
\beq
\left({d^2 \over dt^2} + \omega^2(t)\right)\psi(t) = 0
\label{dyp}
\eeq
where
\beq
\omega^2(t) = m_a^2(t) + {3 \over 16 t^2}~~~\ .
\label{ega}
\eeq
For $t > t_1$, we have ${d \over dt} \ln{\omega} << \omega \simeq m_a$.
That regime is characterized by the adiabatic invariant
$\psi_0^2(t)\omega(t)$, where $\psi_0(t)$ is the changing
oscillation amplitude of $\psi(t)$.  We have therefore
\beq
\psi(t) \simeq {C \over \sqrt{m_a(t)}}
\cos \left(\int^t dt^\prime \omega(t^\prime) \right)~~~~~\ ,
\label{adi}
\eeq
where $C$ is a constant.  Hence
\beq
a(t) = A(t) cos \left(\int^t dt^\prime \omega(t^\prime) \right)~~~,
\label{axt}
\eeq
with
\beq
A(t) = {C \over \sqrt{m_a(t)}} {1 \over t^{3 \over 4}}~~~\ .
\label{oam}
\eeq
Hence, during the adiabatic regime, 
\beq
A^2(t) m_a(t) \propto t^{-{3 \over 2}} \propto R(t)^{-3}~~~\ .
\label{amt}
\eeq
The zero momentum mode of the axion field has energy density
$\rho_a = {1 \over 2} m_a^2 A^2$, and describes a coherent state
of axions at rest with number density $n_a = {1 \over 2} m_a A^2$.
Eq. (\ref{amt}) states therefore that the number of zero momentum
axions per co-moving volume is conserved.  The result holds as 
long as the changes in the axion mass are adiabatic.

We estimate the number density of axions in the zero momentum mode
at late times by saying that the axion field has a random initial
value $a(t_1) = f_a \alpha_1$ and evolves according to Eqs. (\ref{axt},
\ref{oam}) for $t > t_1$.  $\alpha_1$ is called the 'initial misalignment
angle'. Since the effective potential for $a$ is periodic with period 
$2 \pi f_a$, the relevant range of $\alpha_1$ values is $-\pi$ to
$+\pi$.  The number density of zero momentum axions at time $t_1$ 
is then \cite{Pres83,Abb83,Dine83}
\beq
n_a^{\rm vac,0}(t_1) \sim {1 \over 2} m_a(t_1) (a(t_1))^2
= {f_a^2 \over 2 t_1} (\alpha_1)^2
\label{na1}
\eeq
where we used Eq. (\ref{t_1}).  We will use Eq. (\ref{na1}) 
in Section 4 to estimate the zero mode contribution to the
cosmological energy density of cold axions.

A more precise treatment would solve Eq.~(\ref{zeq}) for $t \sim t_1$,
e.g. by numerical integration, to obtain the exact interpolation between
the sudden ($t < t_1$) and adiabatic ($t > t_1$) regimes.  An additional
improvement is to solve Eq.~(\ref{zeq}) without linearizing the sine
function, thus allowing large values of $\alpha_1$.  Although these
improvements are desirable, they would still leave the number of axions
unknown in case 1 because the initial misalignment angle $\alpha_1$ is
unknown.  In case 2, the zero mode contribution to the axion number
density is also given by Eq.~(\ref{na1}) but the misalignment angle 
$\alpha_1$ varies randomly from one horizon to the next.

\subsection{Non zero modes}

In case 2, where there is no inflation after the PQ phase transition,
the axion field is spatially varying.  Axion strings are present
as topological defects, and non zero momentum modes of the axion
field are excited.  In this subsection, we consider a region of the
universe which happens to be free of strings.  Strings will be added
in the next subsection.

The axion field satisfies Eq. (\ref{eom}).  We neglect the axion
mass till $t \sim t_1$.  The solution of Eq.~(\ref{eom}) is a linear
superposition of eigenmodes with definite co-moving wavevector $\vec k$:
\beq
a(\vec x, t) = \int d^3 k~~a(\vec k, t)~e^{i\vec{k} \cdot \vec{x}}
\label{fou}
\eeq
where the $a(\vec k, t)$ satisfy:
\begin{equation}
\left( \partial_t^2 + {3\over 2t} \partial_t +
{k^2\over R^2} \right) a(\vec k, t) = 0\ .
\label{eqk}
\eeq
Eqs.~(\ref{met}) and (\ref{fou}) imply that the wavelength $\lambda (t)
= {2\pi\over k} R(t)$ of each mode is stretched by the Hubble
expansion.  There are two qualitatively different regimes in the
evolution of a mode, depending on whether its wavelength is outside
$(\lambda (t) > t)$ or inside $(\lambda (t) < t)$ the horizon.

For $\lambda (t) \gg t$, only the first two terms in Eq.~(\ref{eqk})
are important and the most general solution is:
\beq
a(\vec k, t) = a_0 (\vec k) + a_{-1/2} (\vec k) t^{-1/2}\ .
\label{abf}
\eeq
Thus, for wavelengths larger than the horizon, each mode goes to
a constant; the axion field is so-called ``frozen by causality''.

For $\lambda (t) \ll t$, let
$a(\vec k, t) = R^{-{3 \over 2}}(t) \psi(\vec k,t)$. Eq.~(\ref{eqk})
becomes
\beq
\left( \partial_t^2 + \omega^2(t) \right) \psi(\vec k, t) = 0 \,
\label{har}
\eeq
where
\beq
\omega^2(t) = {k^2\over R^2 (t)} + {3\over 16t^2} \simeq
{k^2\over R^2(t)}\ .
\label{nom}
\eeq
Since ${d \over dt} \ln(\omega) \ll \omega$, this regime is again
characterized by the adiabatic invariant $\psi_0^2(\vec k,t)\omega(t)$,
where $\psi_0(\vec k,t)$  is the oscillation amplitude of $\psi(\vec
k,t)$.  Hence the most general solution is:
\beq
a (\vec k, t) = {C \over R(t)} \cos \left( \int^t dt^\prime \omega
(t^\prime) \right)~~~~~\ ,
\label{ads}
\eeq
where $C$ is a constant. The energy density and the number density
behave respectively as $\rho_{a,\vec k} \sim {C^2 w^2\over R^2(t)}
\propto {1\over R^4 (t)}$ and
$n_{a,\vec k} \sim {1\over \omega} \rho_{a,\vec k}
\propto {1\over R^3 (t)},$ indicating that the number of axions in
each mode is conserved.  This is as expected because the expansion
of the universe is adiabatic for modes with $\lambda (t) t \ll 1$.

Let us call ${dn_a\over dw} (\omega, t)$ the number density, in
physical and frequency space, of axions with wavelength
$\lambda=\frac{2\pi}{\omega}$, for $\omega > t^{-1}$.  The axion
number density in physical space is thus:
\begin{equation}
n_a (t) = \int_{t^{-1}} d\omega~{dn_a\over dw} (\omega, t)\ ,
\label{3.8}
\eeq
whereas the axion energy density is:
\beq
\rho_a (t) = \int_{t^{-1}} d\omega {dn_a\over d\omega} (\omega, t)
\omega\ .
\label{3.9}
\eeq
Under the Hubble expansion axion energies redshift according to
$\omega^\prime = \omega {R\over R^\prime}$, and volume elements expand
according to
$\Delta V^\prime = \Delta V\left({R^\prime\over R}\right)^3$, whereas
the number of axions is conserved mode by mode. Hence
\beq
{dn_a\over d\omega} (\omega, t) = \left({R^\prime\over R}\right)^2
{dn_a\over d\omega} (\omega {R\over R^\prime}, t^\prime)\ .
\label{3.10}
\eeq
Moreover, the size of ${dn_a\over d\omega}$ for $\omega \sim {1\over t}$
is determined in order of magnitude by the fact that the axion field
typically varies by $v_a = N f_a$ from one horizon to the next.  Thus:
\beq
\left.\omega {dn_a\over d\omega} (\omega, t) \Delta\omega \right|_{\omega
\sim\Delta\omega\sim \frac{1}{t}}
\sim {dn_a\over d\omega}
\left({1\over t}, t\right) \left({1\over
t}\right)^2 \sim {1\over 2} (\vec\nabla a)^2 \sim
{1\over 2} {N^2 f_a^2\over t^2}\ .
\label{3.11}
\eeq
From Eqs.~(\ref{3.10}) and (\ref{3.11}), and $R \propto \sqrt{t}$,
we have \cite{Chang99}
\beq
{dn_a\over d\omega} (\omega, t) \sim {N^2 f_a^2\over 2t^2\omega^2}\ .
\label{3.12}
\eeq
Eq.~(\ref{3.12}) holds until the moment the axion acquires mass during the
QCD phase transition.

\subsection{Strings}

In case 2 axion strings are present as topological defects in the
axion field from the PQ to the QCD phase transitions \cite{Vil82b}.
The energy per unit length of an axion string is 
\begin{equation}
\mu = \pi v_a^2 \ln(v_a L)~~~~\ .
\label{strten}
\end{equation}
$L$ is an infra-red cutoff, which in practice equals the distance 
to the nearest neighbor string.  Because they are strongly coupled 
to the axion field, the strings decay very efficiently into axions.  
We will see that practically all axions produced by string decay 
are non-relativistic after $t_1$.  Because each such axion contributes
$m_a$ to the present energy density, it is important to evaluate the 
{\it number} density of axions emitted in string decay.  This is our 
main goal in this subsection.

At a given time $t$, there is at least on the order of one string
per horizon.  Indeed the axion field is completely uncorrelated
over distances larger than $t$.  Hence there is non-zero probability
that the random values of $a(\vec{x},t)$ wander from zero to $2 \pi v_a$
along a closed path in physical space if that closed path has size larger
than $t$.  When this is the case, a string perforates the surface
subtended by the closed path.

At first, the strings are stuck in the primordial plasma and are
stretched by the Hubble expansion.  During that time, because
$R(t) \propto \sqrt{t}$, the density of strings grows to be much
larger than one per horizon.  However expansion dilutes the plasma
and at some point the strings become unstuck.  The temperature at
which strings start to move freely is of order
\cite{Har87}
\beq
T_* \sim 2~10^7 \mbox{GeV}
\left({f_a \over 10^{12} \mbox{GeV}}\right)^2~~~\ .
\label{tst}
\eeq
Below $T_*$, there is a network of axion strings moving at relativistic
speeds.  Axions are radiated very efficiently by collapsing string loops 
and by oscillating wiggles on long strings.  By definition, long strings
stretch across the horizon.  They move and intersect one another.  When
strings intersect, there is a high probability of reconnection, i.e. of
rerouting of the topological flux \cite{Shel87}.  Because of such
'intercommuting', long strings produce loops which then collapse
freely.  In view of this efficient decay mechanism, the average 
density of long strings is expected to be of order the minimum 
consistent with causality, namely one long string per horizon.  
Hence the energy density in long strings:
\beq
\rho_{\rm str}(t) = \xi {\tau \over t^2}  \simeq
\xi \pi {(f_a N)^2 \over t^2} \ln (v_a t)~~\ ,
\label{std}
\eeq
where $\xi$ is a parameter of order one.

The equations governing the number density $n_a^{\rm str}(t)$ of axions
radiated by axion strings are \cite{Har87}
\beq
{d\rho_{\rm str}\over dt} = -2 H\rho_{\rm str} -
{d\rho_{{\rm str}\rightarrow a}\over dt}
\label{Ha1}
\eeq
and
\beq
{dn_a^{\rm str}\over dt} = - 3H n_a^{\rm str} + {1\over \omega (t)}
{d\rho_{{\rm str}\rightarrow a}\over dt}
\label{Ha2}
\eeq
where $\omega (t)$ is defined by:
\beq
{1\over \omega (t)} = {1\over {d\rho_{{\rm str}\rightarrow a}\over dt}}
\int {dk \over k} {d\rho_{{\rm str}\rightarrow a}\over dt~dk}\ .
\label{rom}
\eeq
$k$ is axion momentum magnitude. ${d\rho_{{\rm str}\rightarrow a}\over
dt}(t)$
is the rate at which energy density gets converted from strings to axions
at time $t$, and ${d\rho_{{\rm str}\rightarrow a}\over dt~dk}(t,k)$ is the
spectrum of the axions produced.  $\omega(t)$ is therefore the average
energy of axions radiated in string decay processes at time $t$.  The
term $-2 H \rho_{\rm str} = + H \rho_{\rm str} - 3 H \rho_{\rm str}$ in
Eq. (\ref{Ha1}) takes account of the fact that the Hubble expansion both
stretches $(+H\rho_{\rm str})$ and dilutes $(-3H\rho_{\rm str})$ long
strings.  Integrating Eqs. (\ref{std} - \ref{Ha2}), setting
$H = {1 \over 2t}$, and neglecting terms of order one versus terms of
order $\ln (v_a t)$, one obtains
\beq
n_a^{\rm str}(t) \simeq
{\xi \pi f_a^2 N^2 \over t^{3 \over 2}} \int_{t_{\rm PQ}}^t
dt^\prime~{\ln(v_a t^\prime) \over 
t^{\prime {3 \over 2}} \omega(t^\prime)}~~\ ,
\label{nsa}
\eeq
where $t_{\rm PQ}$ is the time of the PQ transition.

To obtain $n_a^{\rm str}(t)$ we need to know $\omega(t)$, the average
energy of axions radiated at time $t$.  If $\omega(t)$ is large, the
number of radiated axions is small, and vice-versa.  Axions are radiated
by wiggles on long strings and by collapsing string loops.  Consider a
process which starts at $t_{\rm in}$ and ends at $t_{\rm fin}$, and which
converts an amount of energy $E$ from string to axions.  $t_{\rm in}$ and
$t_{\rm fin}$ are both taken to be of order $t$.  It is useful to define
the quantity
\cite{Hagm91}
\beq
N_{\rm ax}(t) \equiv \int dk {dE \over dk}(t) {1 \over k}~~\ .
\label{Nax}
\eeq
where $k$ is wavevector, and ${dE \over dk}(t)$ is the wavevector
spectrum of the $a$ field.  At the start ($t = t_{\rm in}$), only 
string constributes to the integral in Eq.~(\ref{Nax}).  At the 
end ($t = t_{\rm fin}$), only axions contribute.  In between, both 
axions and string contribute.  The number of axions radiated is
$N_a = N_{\rm ax}(t_{\rm fin})$, and their average energy is 
$\omega = {E \over N_a}$.  The energy stored in string has 
spectrum ${dE \over dk} \propto {1 \over k}$ for 
$k_{\rm min} < k < k_{\rm max}$ where $k_{\rm max}$ is of order $v_a$ 
and $k_{\rm min}$ of order ${2 \pi \over L} \sim {2 \pi \over t}$.  
If $\ell \equiv {E \over \mu}$ is the length of string converted to 
axions, we have 
\beq
N_{\rm ax}(t_{\rm in}) = 
{E \over \ln(t v_a) k_{\rm min}}~~~\ .
\label{Naxin}
\eeq
Hence
\beq
{1 \over \omega} = {r \over \ln(v_a t) k_{\rm min}}
\label{ome}
\eeq
where $r$ is the relative change in $N_{\rm ax}(t)$ during the process
in question:
\beq
r \equiv {N_{\rm ax}(t_{\rm fin}) \over N_{\rm ax}(t_{\rm in})}~~\ .
\label{r}
\eeq
$k_{\rm min}$ is of order ${2\pi \over L}$ where $L$ is the loop
size in the case of collapsing loops, and the wiggle wavelength in
the case of bent strings. $L$ is at most of order $t$ but may be
substantially smaller than that if the string network has a lot of
small scale structure.  To parametrize our ignorance in this matter,
we define $\chi$ such that the suitably averaged
$k_{\rm min} \equiv \chi {2 \pi \over t}$. Combining Eqs. (\ref{nsa}) 
and (\ref{ome}) we find:
\beq
n_a^{\rm str}(t) \simeq 
{\xi \bar{r} N^2 \over \chi} {f_a^2 \over t}~~\ ,
\label{nst}
\eeq
where $\bar{r}$ is the weighted average of $r$ over the various 
processes that convert string to axions.  One can show \cite{Hagm01} 
that the population of axions that were radiated between $t_{\rm PQ}$ 
and $t$ have spectrum ${d n_a \over dk} \propto {1 \over k^2}$ for
${1 \over t} \ltwid k \ltwid {1 \over \sqrt{t t_{\rm PQ}}}$,
irrespective of the shape of ${d\rho_{{\rm str}\rightarrow a}\over
dt~dk}$, provided $t >> t_{\rm PQ}$.

At time $t_1$, each string becomes the edge of $N$ domain walls, and
the process of axion radiation by strings stops.  Since their momenta
are of order $t_1^{-1}$ at time $t_1$, the axions radiated by strings
become non-relativistic soon after they acquire mass.  We discuss the
string decay contribution to the axion energy density in Section 4.
For the moment we turn our attention to the domain walls which appear
at $t_1$ in case 2.

\section{The domain wall problem}

Axion models have an exact, spontaneously broken, discrete $Z(N)$
symmetry.  $Z(N)$ is the subgroup of $U_{\rm PQ}(1)$ which does 
not get broken by non-perturbative QCD effects \cite{Sik82}.  The
spontaneously broken $Z(N)$ symmetry implies a $N$ fold degeneracy 
of the vacuum.  The $N$ vacua are at equidistant points on the circle 
at the bottom of the 'Mexican hat' potential for the Peccei-Quinn field 
$\phi$.  An axion domain wall is the minimum energy field configuration 
which interpolates between neighboring vacua.  Note that there are axion 
domain walls even when $N = 1$.  In this case both sides of the domain 
wall are in the same vacuum (indeed there is only one vacuum) but the 
interpolating field configuration winds around the bottom of the Mexican 
hat potential once.  The properties of walls in $N = 1$ models are for
most purposes identical to those of walls in $N \geq 2$ models.  The
only seemingly important difference is that the walls of $N = 1$ models 
are quantum mechanically unstable.  Even so, their decay rate per unit 
surface and time is exponentially small and negligible in the context 
of our discussion.

When the axion mass turns on, at time $t_1$, each axion string becomes 
the edge of $N$ domain walls.  The domain walls produce a cosmological
disaster unless there is inflation after the PQ phase transition (case 1), 
or unless $N = 1$.  Indeed, let's consider the implications of case 2 if
$N \geq 2$. Since there are two or more exactly degenerate vacua and they 
have identical properties, the vacua chosen at points outside each other's
causal horizon are independent of one another.  Hence there is at least 
on the order of one domain wall per causal horizon at any given time.  
In case 2, the size of the causal horizon is of order $t$, the age of 
the universe.  Thus the energy density in domain walls
\beq
\rho_{\rm w} (t) \gtwid {\sigma \over t}
\label{dwd}
\eeq
where $\sigma$ is the wall energy per unit surface, given by 
\cite{Sik82,Hua85}
\begin{equation}
\sigma \simeq 9 f_a^2 m_a \simeq 
5.5 \cdot 10^{10}~{\rm GeV}^3 
\left({f_a \over 10^{12}~{\rm GeV}}\right)~~~~\ .
\label{wallen}
\end{equation}
The energy density in axion domain walls today ($t_0 \simeq$ 14 Gyr) 
\beq
\rho_{\rm w} (t_0) \gtwid {\sigma \over t_0} \simeq
2 \cdot 10^{-14} {{\rm gr} \over {\rm cm}^3}
\left({f_a \over 10^{12} {\rm GeV}}\right)
\label{dwt}
\eeq
would exceed by many orders of magnitude the critical energy density, 
of order $10^{-29}$ gr/cm$^3$, for closing the universe.  This would 
be grossly inconsistent with observation.  Let's see what would happen.

Let $t_{\rm w}$ be the age of the universe when the domain 
walls start to dominate the energy density.  The condition 
$H^2 \sim {8\pi G\over 3} \rho_{\rm w}$ and Eq. (\ref{dwd}) imply
\beq
t_{\rm w} \ltwid {3\over 32 \pi G\sigma} \simeq 53~{\rm sec}
\left({10^{12} {\rm GeV} \over f_a}\right)\ .
\label{twa}
\eeq
Domain walls are gravitationally repulsive \cite{Vil81,Ips84,Vil83}.
They accelerate away from each other with acceleration $2 \pi G \sigma$ 
and, after a time of order $(2 \pi G \sigma)^{-1}$, recede at the speed 
of light.  By averaging over volumes containing many cells separated by 
walls, the equation of state of a wall dominated universe is seen to be
\beq
p_{\rm w} = -{2\over 3} \rho_{\rm w}~~~\ .
\label{eqs}
\eeq
Conservation of energy
\beq
d(\rho_{\rm w} R^3) = - p_{\rm w} d(R^3)~~\ ,
\label{enc}
\eeq
where $R$ is the scale factor, then implies
$\rho_{\rm w} \propto {1 \over R}$.  This scaling law and the
Friedmann equation
\beq
H^2 = \left({\dot{R} \over R}\right)^2 = {8\pi G\over 3} \rho_{\rm w}
\label{fdw}
\eeq
imply that a domain wall dominated univers expands according to
\beq
R \propto t^2 ~~\ .
\label{rt2}
\eeq
The domain wall dominated universe has an accelerated expansion.  One
may be tempted to attribute the present acceleration of the expansion
of the universe \cite{Perl99,Ries98} to domain walls.  However, a domain
wall dominated universe is far less homogenous than ours.  It would be
divided into cells separated by rapidly expanding walls.  Inside each
cell, concentrated near the cell's center, would be a clump of matter
and radiation with total energy of order
\beq
M \sim \rho (t_{\rm w}) t_{\rm w}^3 \sim 10^{11} M_\odot
\left({m_a \over {\rm eV}}\right)\ .
\label{7.3.4}
\eeq
One cannot identify these clumps with galaxies because neighboring
clumps are receding from one another at close the speed of light.

There are three solutions to the axion domain wall problem.  The first
solution is to have inflation with reheating temperature less than the 
PQ phase transition temperature, i.e. postulate case 1.  In case 1 the
axion field is homogenized by inflation and there are no strings or 
domain walls, and hence no domain wall problem.  The second solution 
is to postulate $N = 1$.  The third solution is to postulate a small
explicit breaking of the $Z(N)$ symmetry. The viability of the second
and third solutions is less obvious.  We discuss them in succession.

\subsection{$N = 1$}

The above arguments, showing the existence of a domain wall problem,
are valid only when the vacuum is multiply degenerate.  They do not 
apply to the $N = 1$ case.  On the other hand, since $N = 1$ models
contain domain walls too, it is not immediately clear that they are 
free of difficulties.  However, $N = 1$ is a solution \cite{Vil82b,
Laza82,Sik83a}
as we now discuss.  

In the circumstances under consideration (case 2) axion strings are
present in the early universe from the time of the PQ phase transition
to that of the QCD phase transition.  At temperature $T_1$ each string
becomes the boundary of a single domain wall.  To see what the network 
of walls bounded by string looks like, a cross-section of a finite but
statistically significant volume of the universe near time $t_1$ was
simulated in refs. \cite{Sik83a,Chang99}.  The simulation shows that 
there are no infinite domain walls which are not cut up by any string.  
The reason for this is easily understood.  An extended domain wall has 
some probability to be cut up by string in each successive horizon it
traverses.  The probability that no string is encountered after traveling
a distance $l$ along the wall decreases exponentially with $l$.

The question now is: what happens to the network of walls bounded 
by strings.  The walls are transparent to the thermalized particles 
in the primordial soup, whose typical momentum is of order one GeV, 
but the walls have a large reflection coefficient for non-relativistic
axions such as the cold axions that were produced by vacuum realignment 
and wall decay \cite{Hua85}.  The drag on the motion of the walls which
results from their reflecting cold axions is important at time $t_1$ but 
turns off soon afterwards as the cold axions are diluted by the expansion 
of the universe \cite{Chang99}.  The wall energy per unit surface
$\sigma(t) \simeq 8 m_a(t) f_a^2$ is time dependent.  [Eq.~(\ref{wallen}) 
is for zero temperature.]  The string at the boundary of a wall is embedded 
into the wall. Hence its infra-red cutoff $L$, in the sense of 
Eq.~(\ref{strten}), is of order the wall thickness $m_a^{-1}$ \cite{Sik82}.  
The energy per unit length of such string is therefore
\begin{equation}
\mu \simeq \pi f_a^2 \ln (f_a/m_a)\ .
\label{3.29}
\end{equation}
The surface energy $E_\sigma$ of a typical (size $\sim t_1$) piece
of wall bounded by string is $\sigma(t) t_1^2$ whereas the energy in
the boundary is $E_\mu \sim \mu t_1$.  There is a critical time $t_2$
when the ratio
\begin{equation}
\frac{E_\sigma(t)}{E_\mu}\sim \frac{8 m_a(t)t_1}{\pi \ln (f_a/m_a)}
\label{3.30}
\end{equation}
is of order one. Using Eqs.~(\ref{mat}) and (\ref{t1}), one estimates:
\begin{equation}
t_2 \simeq  10^{-6} \mbox{sec}
\left({f_a\over 10^{12} \mbox{GeV}}\right)^{1/3}
\label{3.35n}
\end{equation}
\begin{equation}
T_2 \simeq 600 \mbox{MeV}
\left({10^{12} \mbox{GeV}\over f_a}\right)^{1/6}.
\label{3.36n}
\end{equation}
After $t_2$ the dynamics of the walls bounded by string is dominated by
the energy in the walls whereas before $t_2$ it is dominated by the energy
in the strings. A string attached to a wall is pulled by the wall's
tension. For a straight string and flat wall, the acceleration is:
\begin{eqnarray}
a_s(t) = {\sigma(t)\over\mu}  \simeq \frac{8 m_a(t)}{\pi \ln ( f_a/m_a)}
\simeq \frac{m_a(t)}{23} \simeq \frac{1}{t_1} \frac{m_a(t)}{m_a(t_2)}.
\label{3.31}
\end{eqnarray}
Therefore, after $t_2$, each string typically accelerates to relativistic
speeds, in the direction of the wall to which it is attached, in less than
a Hubble time.  The string will then unzip the wall, releasing the stored
energy in the form of barely relativistic axions.  We estimate in section 
4.3 how much walls bounded by string contribute to the present cosmological 
axion energy density.

A very small portion of the domain wall energy density is in walls
which are not bounded by string, and which form closed surfaces such 
as spheres or torii \cite{Sik83a,Chang99}.  Such closed walls do not 
decay by the process just described.  Instead, they oscillate and emit
gravitational waves \cite{Vil82b}.  Using the quadrupole formula, we 
may estimate the gravitational wave power emitted by a closed wall of 
size $\ell$ oscillating with frequency $\omega \sim \ell^{-1}$:
\begin{equation}
P \sim - {d(\sigma\ell^2)\over dt} \sim G(\sigma\ell^4)^2 \omega^6
\sim G\sigma^2\ell^2\ .
\label{3.32}
\end{equation}
Eq.~(\ref{3.32}) implies the lifetime:
\begin{equation}
\tau_{\rm grav} \sim (G\sigma)^{-1} \simeq 2 \cdot 10^3~\mbox{sec}~
\left({10^{12}~{\rm GeV} \over f_a}\right)\ ,
\label{3.33}
\end{equation}
independently of size.

\subsection{A small breaking of PQ symmetry}

A third solution to the domain wall problem is to postulate a small
explicit breaking of the $Z(N)$ symmetry, and hence of PQ symmetry
\cite{Sik82}.  The symmetry breaking must lift completely the degeneracy
of the vacuum and be large enough that the unique true vacuum takes over
before the walls dominate the energy density. On the other hand, it must
be small enough that the PQ mechanism still works.  This solution does 
not appear very attactive and we will see below that there is little 
room in parameter space for it to occur, but it is a logical possibility.

To get rid of the walls, we add to the RHS of Eq.~(\ref{lphi}) a 
tiny $U_{\rm PQ}(1)$ breaking term which lifts the vacuum degeneracy
completely, e.g.:
\begin{equation}
\delta V = - \xi (\phi e^{-i \delta} + h.c.)~~~\ .
\label {5.3}
\end{equation}
To add such a term to an axion model by hand seems rather
unnatural.  However it is conceivable that a small $U_{\rm PQ}(1)$
breaking term is in fact a natural property of the ultimate theory.  
This would be the case, for example, if the low energy effective 
theory at some energy scale has an automatic PQ symmetry which is 
broken in the full theory.  Be that as it may, Eq. (\ref{5.3}) yields 
a small correction to the effective potential for the axion field:
\begin{equation}
\delta V_a = - 2 v_a\xi~\cos \left({a\over v_a} - \delta\right)\ .
\label{5.4}
\end{equation}
The unique true vacuum is the one for which $\mid \delta -
{a\over v_a}\mid$ is smallest.  Its energy density is lowered by an
amount of order $\xi v_a$ relative to the other now quasi-vacua.  As a
result, the walls at the boundary of a region in the true vacuum are
subjected to an outward pressure of order $\xi v_a$.  Since the walls
are typically a distance $t$ apart, the volume energy $\xi~v_a~t^3$
associated with the lifting of the vacuum degeneracy grows more rapidly
than the energy $\sigma t^2$ in the walls.  At a time
$\tau \sim {\sigma\over \xi v_a}$, the pressure favoring the
true vacuum starts to dominate the wall dynamics and the true 
vacuum takes over, i.e. the walls disappear.  The energy stored 
in the network of walls and strings decays into gravitational waves 
\cite{Chang99}.  The true vacuum must take over before the walls 
dominate the energy density.  Using Eq.~(\ref{twa}), we obtain:
\begin{equation}
\tau \sim {\sigma\over \xi v_a} \ltwid 10^2~\mbox{sec}~
\left({10^{12}~{\rm GeV} \over f_a}\right)\ .
\label{5.5}
\end{equation}
On the other hand $\xi$ is bounded from above by the requirement
that $\delta V$ does not upset the PQ mechanism.  $\delta V$ shifts the
minimum of the effective potential for the axion field, inducing a
$\bar\theta \sim {\xi\over m_a^2 f_a}$.  The requirement that
$\bar\theta < 10^{-10}$ implies:
\begin{equation}
\tau \gtwid {10~\mbox{sec}\over N} 
\left({f_a \over 10^{12}~{\rm GeV}}\right)~~~\ .
\label{5.7}
\end{equation}
Eqs.~(\ref{5.5}) and (\ref{5.7}) indicate that there is little 
room in parameter space for this third solution to the axion 
domain wall problem, but it is not ruled out.

\section{Cold axions}

Our main purpose in this Section is to estimate the cosmological 
energy density in cold axions, and estimate their velocity dispersion.  
The discussion in the previous two sections identified the following
sources of cold axions:\\
\indent- vacuum realignment\\
\indent\hspace{.7cm}- zero momentum mode\\
\indent\hspace{.7cm}- higher momentum modes\\
\indent- axion string decay\\
\indent- axion domain wall decay.\\
In case 1, only the contribution from vacuum realignment with
zero momentum is present.  In case 2, all sources are present.

\subsection{Vacuum realignment}

{\it Zero momentum mode}

\vspace{0.2in}

Let us start with case 1 which is easiest to analyze.  The axion
number density at time $t_1$ is given by Eq. (\ref{na1}) in terms
of the initial vacuum misalignment angle $\alpha_1$.  In case 1,
where the universe was homogenized by inflation after the PQ
phase transition, $\alpha_1$ has the same value everywhere.
We saw in Section 2 that, barring any sudden changes in the
axion mass during the chiral symmetry breaking phase transition,
the number of axions is an adiabatic invariant after $t_1$.  Hence
the axion density from the vacuum realignment zero mode at a later
time $t$
\begin{equation}
n_a^{{\rm vac},0}(t) \sim {f_a^2 \over 2 t_1}
(\alpha_1)^2 \left({R_1 \over R}\right)^3~~~~~~~~~~~~~{\rm for~case~1}
\label{numdent}
\end{equation}
where ${R_1 \over R}$ is the ratio of scale factors between
$t_1$ and $t$.  Thus we find
\begin{equation}
\rho_a^{{\rm vac},0} (t_0) \sim {m_a f_a^2 \over 2 t_1}
(\alpha_1)^2 \left({R_1 \over R_0}\right)^3~~~~~~~~~{\rm for~case~1.}
\label{enden0}
\end{equation}
for the axion energy density today.

In case 2, the initial misalignment angle $\alpha_1$ is different
from one QCD horizon to the next.  Since the average of $(\alpha_1)^2$
over many QCD horizons is of order one, we have
\begin{equation}
\rho_a^{{\rm vac},0} (t_0) \sim {m_a f_a^2 \over 2 t_1}
\left({R_1 \over R_0}\right)^3~~~~~~~~~~~~~~~{\rm for~case~2.}
\label{ed20}
\end{equation}
However in case 2, there are additional contributions.

\vspace{0.2in}

{\it Higher momentum modes}

\vspace{0.2in}

In case 2, the axion field is not constant even within each
horizon volume.  It has wiggles inherited from earlier epochs when
the horizon was smaller and the axion field was inhomogeneous on
correspondingly shorter scales.  The associated density of axions 
in physical and frequency space is given in Eq.~(\ref{3.12}).
Integrating over $\omega > t_1^{-1}$, we find the contribution
from vacuum realignment involving higher momentum modes
\begin{equation}
n_a^{{\rm vac},1}(t_1) \sim {N^2 f_a^2 \over 2 t_1}~~~~~
~~~~~~~~~~~~{\rm for~case~2.}
\label{ndv1}
\end{equation}
Almost all these axions are non-relativistic after $t_1$.  Hence
\begin{equation}
\rho_a^{{\rm vac},1} (t_0) \sim {m_a N^2 f_a^2 \over 2 t_1}
\left({R_1 \over R_0}\right)^3~~~~~{\rm for~case~2.}
\label{edv10}
\end{equation}
Note that, except for the factor $N^2$, $\rho_a^{{\rm vac},1}$
and $\rho_a^{{\rm vac},0}$ are of the same order of magnitude.

\subsection{String decay}

The evolution of axion strings between the PQ and QCD phase
transitions was discussed in section 2.3 .  The number density 
of axions emitted in string decay is given by Eq.~(\ref{nst}) in 
terms of quantities $\xi,~\bar{r}$, and $\chi$, which we introduced 
to parametrize our ignorance of various aspects of string evolution 
and decay.

Because their spectrum ${dn_a \over dk} \propto {1 \over k^2}$ over 
the range ${1 \over t} \ltwid k \ltwid {1 \over \sqrt{t t_{\rm PQ}}}$,
the bulk of axions emitted in string decay have momenta of order
$t_1^{-1}$ at time $t_1$, and become non-relativistic soon after they
acquire mass.  Therefore, the string decay contribution to the axion
energy density today is
\begin{equation}
\rho_a^{\rm str}(t_0) = m_a n_a^{\rm str}(t_1)
\left({R_1 \over R_0}\right)^3 \simeq
m_a {\xi \bar{r} \over \chi} {N^2 f_a^2 \over t_1}
\left({R_1 \over R_0}\right)^3~~~~~~~~\ .
\label{sed0}
\end{equation}
We now discuss the factors on the RHS of Eq.~(\ref{sed0})
which are specific to the string decay contribution.

$\xi$~~~This parameter determines the density of the string 
network, Eq.~(\ref{std}), with $\xi =1$ corresponding to a 
density of one long string per horizon.  In ref. \cite{Hagm91} 
it was argued that $\xi \simeq 1$ because global strings can 
decay efficiently into axions, and therefore the number density 
of long strings should be close to the minimum consistent with
causality. In numerical simulations of global string networks 
in an expanding universe \cite{Yamag99}, it was found that 
indeed $\xi\simeq 1$.  So there appears to be good ground
for setting $\xi \simeq 1$.

$\chi$~~~This parameter defines the low wavevector edge of
the ${dn_a \over dk} \propto {1 \over k^2}$ spectrum through
$k_{\rm min} \equiv \chi {2 \pi \over t}$.  $\chi$ and $\xi$
are related since the average interstring distance controls
both.  On dimensional grounds, $\chi \propto \sqrt{\xi}$.  So
the effect of small scale structure in the axion string network
partially cancels out in the RHS of Eq.~(\ref{sed0}).  $\chi$
is expected to be of order one, but the uncertainty on this is
at least a factor two.

$\bar{r}$~~~~~This parameter defines the average energy of the 
axions emitted in string decay, through Eqs.~(\ref{ome},\ref{r}).  
It is the unknown on which most of the debate has focused in the
past.  Two basic scenarios have been put forth, which we call A 
and B.  The question is: what is the spectrum of axions radiated 
by strings?  The main source is closed loops of size $L \sim t$.
Scenario A postulates that a bent string or closed loop oscillates
many times, with period of order $L$, before it has released its
excess energy and that the spectrum of radiated axions is
concentrated near ${2 \pi \over L}$.  In that case one has
$\bar{r} \sim \ln({v_a t_1}) \simeq 67$.  Scenario B
postulates that the bent string or closed loop releases its
excess energy very quickly and that the spectrum of radiated
axions is ${dE \over dk} \propto {1 \over k}$ with a high
frequency cutoff of order $2 \pi v_a$ and a low frequency 
cutoff of order ${2 \pi \over L}$.  In scenario B, the initial 
and final spectra ${dE \over dk}$ of the energy stored in the 
axion field are qualitatively the same and hence $\bar{r} \sim 1$.  
In scenario A, the string decay contribution dominates over the 
vacuum realignment contribution by the factor $\ln(v_a t_1)$, 
whereas in scenario B the contributions from string decay and 
vacuum realignment have the same order of magnitude.

Many authors \cite{Davis85,Vil87,Davis89,Dabh90,Batt94} have argued 
in favor of scenario A, adopting the point of view that global strings 
are similar to local strings and that their coupling to the axion field
can be treated perturbatively. My collaborators and I \cite{Har87,Hagm91}
have argued in support of scenario B, emphasizing that the dynamics of
global strings is dominated by the energy stored in the axion field and
that there is no reason to believe that this energy would behave in 
the same way as the energy stored in the string core. The numerical
simulations of the motion and decay of axion strings in 
refs. \cite{Hagm91,Hagm01} give strong support to scenario B.
These simulations are of oscillating strings with ends held fixed,
of collapsing circular loops, and of collapsing non-circular closed
loops with angular momentum.  Over the range of $\ln(v_a L)$ accessible
with present technology [$2.5 \ltwid \ln(v_a L) \ltwid 5.0$], it was 
found that $r \simeq 0.8$ for closed loops and $r \simeq$ 1.07 for
oscillating strings with ends held fixed.  No dependence of $r$ on
$\ln(v_a L)$ was found for closed loops, and for bent strings with 
ends held fixed $r$ was found to slightly {\it decrease} with increasing
$\ln(v_a L)$, whereas scenario A predicts $r$ to be proportional to
$\ln(v_a L)$.

\subsection{Wall decay}

The final contribution to the cold axion cosmological energy
density in case 2 is from the decay into non-relativistic axions
of axion walls bounded by string.  We assume here that $N = 1$.
Indeed if $N > 1$, the domain wall problem is presumably solved
by introducing a small breaking of PQ symmetry, as described in
section 3.2.  In that case, the axion walls decay predominantly
into gravitational radiation \cite{Chang99}.

Let $t_3$ be the time when the decay effectively takes place and
$\gamma \equiv {\omega^\prime \over m_a(t_3)}$ the average Lorentz 
$\gamma$ factor of the axions produced.  $\omega^\prime$ is their 
average energy.  The density of walls at time $t_1$ was estimated 
to be of order $0.7$ per horizon volume \cite{Chang99}.  Hence the 
average energy density in walls is
\begin{equation}
\rho_{\rm d.w.}(t) 
\sim 0.7~{\sigma(t) \over t_1} \left(\frac{R_1}{R}\right)^3
\sim (0.7) (8) m_a(t)\frac{f_a^2}{t_1} \left(\frac{R_1}{R}\right)^3
\label{3.41n}
\end{equation}
between $t_1$ and $t_3$. We assumed that the energy in walls simply 
scales as $\sigma(t)$. After time $t_3$, the number density of axions 
produced in the decay of walls bounded by strings is of order
\begin{equation}
n^{\rm d.w.}_a(t)\sim \frac{\rho_{\rm d.w.}(t_3)}{\omega^\prime} 
\left(\frac{R_3}{R}\right)^3\sim \frac{6}{\gamma}
\frac{f_a^2}{t_1}\left(\frac{R_1}{R}\right)^3 \, .
\label{3.42n}
\end{equation}
Note that the dependence on $t_3$ drops out of our estimate of
$n^{\rm d.w.}_a$.  In the simulations of the motion and decay
of walls bounded by string in ref. \cite{Chang99} it was found
that $\gamma \simeq 7$ for $\ln({v_a \over m_a}) \sim 4.6$ but
that $\gamma$ increases approximately linearly with
ln$(\sqrt{\lambda} v_a/m_a)$. If this behaviour is extrapolated
all the way to ln$(\sqrt{\lambda} v_a/m_a) \simeq 60$, which is
the value in axion models of interest, then $\gamma \simeq 60$.
In that case the contribution from wall decay is subdominant
relative to those from vacuum realignment and string decay.

\subsection{The cold axion cosmological energy density}
\label{cacedr}

To estimate the cosmological energy density of cold axions in 
case 2, we neglect the contribution from wall decay and assume 
that scenario B is correct for the string contribution.  By
adding the RHS of Eqs. (\ref{ed20}), (\ref{edv10}), and (\ref{sed0}) 
with $N = \bar{r} = \xi = \chi = 1$, we find
\begin{equation}
\rho_a (t_0) \sim 2~{f_a^2\over t_1}
\left({R_1\over R_0}\right)^3 m_a~~~~~{\rm for~case~2.}
\label{3.38}
\end{equation}
Eq. (\ref{enden0}) gives the cold axion cosmological energy density 
in case 1.  To determine the ratio of scale factors $R_1/R_0$, we assume
conservation of entropy from time $t_1$ till the present.  The number 
${\cal N}_1$ of effective thermal degrees of freedom at time $t_1$ is
of order 60.  Keeping in mind that neutrinos decouple before electron-
positron annihilation, one finds
\begin{equation}
\left({R_1\over R_0}\right)^3 \simeq 0.063 \left({T_{\gamma, 0}\over T_1}
\right)^3\ .
\label{3.39}
\end{equation}
Combining Eqs.~(\ref{t1}), (\ref{3.38}), (\ref{enden0}) and
(\ref{3.39}), and dividing by the critical density 
$\rho_c = {3H_0^2\over 8\pi G}$, we find
\begin{eqnarray}
\Omega_a &\sim& 0.15 \left({f_a\over 10^{12}~\mbox{GeV}}\right)^{7/6}
\left({0.7\over h}\right)^2~\alpha_1^2~~~~~~~~~~{\rm
for~case~1}\nonumber\\
&\sim& 0.7 \left({f_a\over 10^{12}~\mbox{GeV}}\right)^{7/6}
\left({0.7\over h}\right)^2~~~~~~~~~~~~~~~~{\rm for~case~2}
\label{3.41}
\end{eqnarray}
where $h$ is defined as usual by $H_0 = h~100 $km/s$\cdot$Mpc.

Eqs.~(\ref{3.41}) are subject to many sources of uncertainty, aside from
the uncertainty about the contribution from string decay.  The axion
energy density may be diluted by the entropy release from heavy particles
which decouple before the QCD epoch but decay afterwards \cite{Stei83, 
Laza87,Laza90}, or by the entropy release associated with a first order
QCD phase transition.  On the other hand, if the QCD phase transition is 
first order \cite{Unr85,Turn85,DeGr86,Hind92}, an abrupt change of the 
axion mass at the transition may increase $\Omega_a$.  A model has been 
put forth \cite{Kapl06} in which the axion decay constant $f_a$ is
time-dependent, the value $f_a(t_1)$ during the QCD phase-transition 
being much smaller than the value $f_a$ today. This yields a 
suppression of the axion cosmological energy density by a factor
$({f_a(t_1) \over f_a})^2$ compared to the usual case.  Finally, it 
has been proposed that the axion density is diluted by 'coherent
deexcitation', i.e. adiabatic level crossing of $m_a(t)$ with the 
mass of some other pseudo-Nambu-Goldstone boson which mixes with the 
axion \cite{Hill88}.

\subsection{Velocity dispersions}

The axions produced by vacuum realignment, string decay and wall
decay all have today extremely small velocity dispersion.  In case 1, 
where the axions are produced in a zero momentum state, the velocity 
dispersion is zero.  (This ignores the small quantum mechanical
fluctuations created during the inflationary epoch, which will be 
discussed in Section 6.)

In case 2, we distinguish two subpopulations of cold axions: pop. I and
pop. II, with the second kind having velocity dispersion typically a
factor $10^3$ to $10^4$ larger than the first.  The pop. I axions are
those produced by vacuum realignment or string decay and which escaped
being hit by moving domain walls.  They have typical momentum
$p_{\rm I}(t_1) \sim \frac{1}{t_1}$ at time $t_1$ because they 
are associated with axion field configurations which are
inhomogeneous on the horizon scale at that time.  Their velocity
dispersion is of order:
\begin{equation}
\beta_{\rm I}(t)\sim \frac{1}{m_at_1}\left(\frac{R_1}{R}\right)
\simeq 3\cdot 10^{-17}
\left(\frac{10^{-5}{\rm eV}}{m_a}\right)^{5/6}\frac{R_0}{R} \ .
\label{3.48}
\end{equation}
The corresponding effective temperature is of order 
$0.5~10^{-34}$ K $({10^{-5} {\rm eV} \over m_a})^{2 \over 3}$ today.
This very cold, indeed!

Pop. II are axions produced in the decay of domain walls and 
axions that were hit by moving domain walls.  Axions produced 
in the decay of domain walls have typical momentum 
$p_{\rm II}(t_3) \sim \gamma m_a(t_3)$ at time $t_3$ when the 
walls effectively decay.  Their velocity dispersion is therefore 
of order:
\begin{equation}
\beta_{\rm II}(t)\sim \gamma \frac{m_a(t_3)}{m_a}\frac{R_3}{R} \simeq
10^{-13}~q~\left(\frac{10^{-5}{\rm eV}}{m_a}\right)^{1/6} \frac{R_0}{R} \,
.
\label{3.49}
\end{equation}
where $q \equiv \gamma \frac{m_a(t_3)}{m_a}\frac{R_3}{R_1}$ parametrizes
our ignorance of the wall decay process.  We expect $q$ to be of order one
but with very large uncertainties.  There is however a lower bound on $q$
which follows from the fact that the time $t_3$ when the walls effectively
decay must be after $t_2$ when the energy density in walls starts to
exceed
the energy density in strings. Using Eq.~(\ref{3.35n}), we have
\begin{equation}
q=\frac{\gamma m_a(t_3)}{m_a}\frac{R_3}{R_1} > \frac{\gamma
m_a(t_2)}{m_a}\frac{R_2}{R_1} \simeq \frac{\gamma}{130}\left(
\frac{10^{-5}{\rm eV}}{m_a}\right)^{2/3} .
\label{5.6n}
\end{equation}
Since computer simulations suggest $\gamma$ is of order 60, pop. II
axions have much larger velocity dispersion than pop. I axions, by a
factor $10^3$ or more.  Whereas pop. II axions are relativistic or near
relativistic at the end of the QCD phase transition, pop. I axions are
definitely non-relativistic at that time since $m_a >> 1/t_1$.  The 
axions which were produced by vacuum realignment or string decay but
were hit by relativistically moving walls at some time between $t_1$ and
$t_3$ should be included in pop. II since they are relativistic just
after getting hit.  The next Section will highlight the differences in
the behaviours of the two populations of cold axions.

The very low velocity dispersion of cold axions, and their 
extremely weak couplings, imply that these particles behave 
as cold collisionless dark matter (CDM).  CDM particles lie 
at all times on a 3-dim. hypersurface in 6-dim. phase-space 
\cite{Ips92,Nata05}.  As a result, CDM forms discrete flows 
and caustics.  The number of discrete flows at a given physical 
location is the number of times the 3-dim. hypersurface covers 
physical space at that location.  At the boundaries between 
regions with differing number of flows, the 3-dim. hypersurface 
is tangent to velocity space.  The dark matter density is very 
large on these surfaces, which are called caustics.  The density 
diverges at the caustics in the limit of zero velocity dispersion.

\section{Axion miniclusters}
\label{am}

If there is no inflation after the PQ phase transition (case 2), the
initial misalignment angle $\alpha_1$ changes by ${\cal O}(1)$ from
one QCD time horizon to the next.  Hence, the fluid of cold axions
produced by vacuum realignment is inhomogeneous with 
${\delta\rho_a\over \rho_a} = {\cal O}(1)$ at the time of the QCD 
phase transition.  As will be shown shortly, the streaming length 
of pop.~I axions is too short for these inhomogeneities to get erased 
by free streaming before the time $t_{\rm eq}$ of equality between 
matter and radiation, when density perturbations start to grow in 
earnest by gravitational instability.  At time $t_{\rm eq}$, the
${\delta\rho_a\over \rho_a} = {\cal O}(1)$ inhomogeneities in the
axion fluid promptly form gravitationally bound objects, called
axion miniclusters \cite{Hog88,Kolb93,Kolb96,Chang99}.  The
properties of axion miniclusters are of concern to experimentalists
attempting the direct detection of dark matter axions on Earth.  
Indeed those experiments would become even more challenging (than 
they are already) if most of the cold axions condense into miniclusters
and the miniclusters withstand tidal disruption afterwards.  Of course,
these issues only arise in case 2.  There are no axion miniclusters in
case 1.

As described above, there are two populations of cold axions,
pop.~I and pop.~II, with velocity dispersions given by Eqs.~(\ref{3.48})
and (\ref{3.49}) respectively.  Both populations are inhomogeneous
at the time of the QCD phase transition.  The free streaming length
from time $t_1$ to $t_{\rm eq}$ is:
\begin{eqnarray}
\ell_{\rm f} =
R(t_{\rm eq}) \int_{t_1}^{t_{\rm eq}} dt {\beta (t)\over R(t)}
\simeq \beta (t_1) \sqrt{t_1 t_{\rm eq}}
\ln \left( {t_{\rm eq}\over t_1}\right) \ .
\label{6.1}
\end{eqnarray}
The time of equality and the corresponding temperature are respectively
$t_{\rm eq} \simeq 2.3~10^{12}$ sec and $T_{\rm eq} \simeq 0.77$ eV.
The free streaming length should be compared with the size
\begin{equation}
\ell_{\rm mc}\sim t_1 {R_{eq}\over R_1} \simeq \sqrt{t_1 t_{\rm eq}}
\simeq
2\cdot 10^{13}{\rm cm} \left(\frac{10^{-5}{\rm eV}}{m_a}\right)^{1/6}
\label{lmc}
\end{equation}
of axion inhomogeneities at $t_{\rm eq}$.  Using Eq.~(\ref{3.48}) we
find for pop.~I:
\begin{equation}
\frac{\ell_{\rm f,I}}{\ell_{\rm mc}} \simeq {1 \over t_1 m_a}
\ln \left({t_{\rm eq}\over t_1}\right) \simeq
2 \cdot 10^{-2} \left(\frac{10^{-5}{\rm eV}}{m_a}\right)^{2/3} .
\label{6.2}
\end{equation}
Hence, in the axion mass range of interest, pop.~I axions
do not homogenize.  At $t_{\rm eq}$ most pop.~I axions condense
into miniclusters.  The typical size of axion miniclusters is 
$\ell_{\rm mc}$ and their typical mass is \cite{Kolb96,Chang99}
\begin{equation}
M_{\rm mc} \sim \eta~\rho_a (t_{\rm eq}) \ell_{\rm mc}^3
\sim \eta~5\cdot 10^{-13} M_\odot
\left(\frac{10^{-5}{\rm eV}}{m_a}\right)^{5/3} ,
\label{6.5}
\end{equation}
where $\eta$ is the fraction of cold axions that are pop. I.  We 
assumed that all pop. I axions condense into miniclusters, and 
used Eq.~(\ref{3.41} - case 2) to estimate $\rho_a (t_{\rm eq})$.

Using Eq.~(\ref{3.49}), we find for pop.~II:
\begin{equation}
\frac{\ell_{\rm f,II}}{\ell_{\rm mc}} \sim
q \ln \left({t_{\rm eq}\over t_3}\right) \simeq 42\, q .
\label{5.5n}
\end{equation}
Using Eq.~(\ref{5.6n}) and assuming the range $\gamma \sim $ 7 to 60,
suggested by the numerical simulations of ref. \cite{Chang99}, we
conclude that pop.~II axions do homogenize and hence that the axion
energy density has a smooth component at $t_{\rm eq}$.

However, pop.~II axions may get gravitationally bound to miniclusters
later on.  It seems rather difficult to model this process reliably.
A discussion is given in ref.~\cite{Chang99}.  It is concluded there
that the accretion of pop.~II axions results in miniclusters which 
have an inner core of pop.~I axions with density of order 
$10^{-18}$ gr/cm$^3$ and a fluffy envelope of pop.~II axions with 
density of order $10^{-25}$ gr/cm$^3$.  

When a minicluster falls onto a galaxy, tidal forces are apt to 
destroy it.  If a minicluster falls through the inner parts of the 
Galaxy ($r < 10$ kpc) where the density is of order $10^{-24}$ gr/cm$^3$, 
its fluffy envelope of pop.~II axions will likely be pulled off immediately.  
This is helpful for direct searches of dark matter axions on Earth since it 
implies that a smooth component of dark matter axions with density of order 
the halo density permeates us whether or not there is inflation after the 
Peccei-Quinn phase transition. Even the central cores of pop.~I axions may
eventually get destroyed. When a minicluster passes by an object of mass 
$M$ with impact parameter $b$ and velocity $v$, the internal energy per 
unit mass $\Delta E$ given to the minicluster by the tidal gravitational 
forces from that object is of order \cite{Hog88}
\begin{equation}
\Delta E \sim {G^2 M^2\ell_{mc}^2\over b^4 \beta^2}
\label{6.9}
\end{equation}
whereas the binding energy per unit mass of the minicluster
$E\sim G~\rho_{mc} \ell_{mc}^2$.  If the minicluster travels a length
$L = \beta t$ through a region where objects of mass $M$ have density
$n$, the relative increase in internal energy is:
\begin{equation}
{\Delta E\over E} \sim {G\rho_M^2 t^2\over \rho_{mc}}\ ,
\label{6.10}
\end{equation}
where $\rho_M = Mn$.  Eq.~(\ref{6.10}) follows from the fact that
$\Delta E$ is dominated by the closest encounter and the latter has
impact parameter $b_{min}$:  $\pi b_{min}^2 n L = 0(1)$.  Note that
${\Delta E\over E}$ is independent of $M$.  A minicluster inner core
which has spent most of its life in the central part of our galaxy 
only barely survived since ${\Delta E\over E} \sim 10^{-2}$ in that
case.  

The direct encounter of a minicluster with Earth would be quite 
rare, happening only every $10^4$ years or so.  The encounter would 
last for about 3 days during which the local axion density would 
increase by a factor of order $10^6$.

\section{Axion isocurvature perturbations}

In this Section we describe the isocurvature perturbations 
\cite{Axen83,Stei83,Lind85,Seck85,Lyth90,Turn91} produced 
if inflation occurs after the Peccei-Quinn phase transition, 
and derive the constraints on axion parameters from the absence 
of isocurvature fluctuations in CMBR observations.

If the reheat temperature after inflation is less than the 
temperature $T_{\rm PQ}$ at which $U_{\rm PQ}(1)$ is restored 
(case 1), the axion field is present during inflation and is 
subject to quantum mechanical fluctuations, just like the inflaton.  
In fact, since the axion field is massless and weakly coupled like 
the inflaton, it has the same fluctuation spectrum 
\cite{Birr82,Vil82c,Lind82,Star82}
\begin{equation}
P_a (k) \equiv \int {d^3 x \over (2 \pi)^3}
<\delta a(\vec{x},t) \delta a(\vec{x}^\prime,t)>
e^{- i \vec{k} \cdot (\vec{x} - \vec{x}^\prime)} =
\left({H_I \over 2 \pi}\right)^2 {2 \pi^2 \over k^3}~~~\ ,
\label{axpow}
\end{equation}
where $H_I$ is the expansion rate during inflation.  As before,
$\vec{x}$ are comoving spatial coordinates.  The axion fluctuations
described by Eq.~(\ref{axpow}) are commonly written in shorthand 
notation as $\delta a = {H_I \over 2 \pi}$.  The fluctuation in 
each axion field mode is ``frozen in" after $R(t)/k$ exceeds the 
horizon length $H_I^{-1}$.

We do not consider here the possibility of fluctuations in the
axion decay constant $f_a$ during inflation.  Such fluctuations 
are discussed in Refs. \cite{Lind91,Lyth92a,Lyth92b}.

At the start of the QCD phase transition, the local value of the axion
field $a(\vec{x},t)$ determines the local number density of cold axions
produced by the vacuum realignment mechanism [see Eq.~(\ref{na1})]~:
\begin{equation} 
n_a(\vec{x}, t_1) = {f_a^2 \over 2 t_1} \alpha(\vec{x},t_1)^2 
\label{locna1} 
\end{equation} 
where $\alpha(\vec{x}, t_1) = a(\vec{x}, t_1)/f_a$ is the misalignment 
angle.  The fluctuations in the axion field produce perturbations in 
the cold axion density
\begin{equation} 
{\delta n_a^{\rm iso} \over n_a} = 
{2 \delta a \over a_1} = {H_I \over \pi f_a \alpha_1} 
\label{deltana}
\end{equation} 
where $a_1 = a(t_1) = f_a \alpha_1$ is the value of the axion 
field at the start of the QCD phase transition, common to our 
entire visible universe.  These perturbations obey  
\begin{equation} 
\delta \rho_a^{\rm iso} (t_1) = 
- \delta \rho_{\rm rad}^{\rm iso}(t_1) 
\label{iso}
\end{equation} 
since the vacuum realignment mechanism converts energy stored in 
the quark-gluon plasma into axion rest mass energy.  In contrast, 
the density perturbations produced by the fluctuations in the inflaton 
field \cite{Mukh81,Hawk82,Star82,Guth82,Bard83} satisfy 
\begin{equation}
{\delta \rho_{\rm matter} \over \rho_{\rm matter}} = 
{3 \over 4} {\delta \rho_{\rm rad} \over \rho_{\rm rad}}~~~\ . 
\label{adiab}
\end{equation} 
Density perturbations that satisfy Eq.~(\ref{adiab}) are called 
``adiabatic", whereas density perturbations that do not satisfy 
Eq.~(\ref{adiab}) are called ``isocurvature".  Isocurvature 
perturbations, such as the density perturbations of Eq.~(\ref{iso}), 
make a different imprint on the cosmic microwave background than do 
adiabatic ones.  The CMBR observations are consistent with pure 
adiabatic perturbations.  This places a constraint on axion models 
if the Peccei-Quinn phase transition occurs before inflation.

Before we derive this constraint, two comments are in order.  The 
first is that, if the Peccei-Quinn transition occurs {\it after}
inflation, axion models still predict isocurvature perturbations 
but not on length scales relevant to CMBR observations.  Indeed 
we saw in the previous section that in this case (case 2) the axion 
field fluctuates by order $f_a$ from one QCD horizon to the next.  
Those fluctuations produce isocurvature perturbations on the scale 
of the QCD horizon, which is much smaller than the length scales 
observed in the CMBR.  Their main phenomenological implication are 
the axion miniclusters which were discussed in Section 5. The second
comment is that, if the Peccei-Quinn phase transition occurs before
inflation (case 1), the density perturbations in the cold axion fluid 
have both adiabatic and isocurvature components.  The adiabatic
perturbations
(${\delta \rho_a^{\rm ad} \over 3 \rho_a} =
{\delta \rho_{\rm rad}^{\rm ad} \over 4 \rho_{\rm rad}}
= {\delta T \over T}$) are produced by the quantum
mechanical fluctuations of the inflaton field during
inflation, whereas the isocurvature perturbations
are produced by the quantum mechanical fluctuations of
the axion field during that same epoch.  The adiabatic 
and axion isocurvature components are uncorrelated.

The upper bound from CMBR observations and large scale structure data
on the fraction of CDM perturbations which are isocurvature is of order
30\% in amplitude (10\% in the power spectrum) \cite{Peir03,Vali03,
Crot03,Belt06,Bean06,Trot06}. Allowing for the possibility that only 
part of the cold dark matter is axions, the bound on isocurvature
perturbations implies
\begin{equation}
{\delta \rho_a^{\rm iso} \over \rho_{\rm CDM}} =
{\delta\rho_a^{\rm iso} \over \rho_a} \cdot {\rho_a \over \rho_{\rm CDM}}
= {H_I \over \pi f_a \alpha_1} {\Omega_a \over \Omega_{\rm CDM}} <
0.3~{\delta \rho_m \over \rho_m}~~~~\ ,
\label{isocon}
\end{equation}
where we used Eq.~(\ref{deltana}). ${\delta \rho_m \over \rho_m}$
is the amplitude of the primordial spectrum of matter perturbations.
It is related to the amplitude low multipole CMBR anisotropies through 
the Sachs-Wolfe effect \cite{Sach67,Peeb82,Abbo84}.  The observations
imply ${\delta \rho_m \over \rho_m} \simeq 4.6~10^{-5}$ \cite{Dode03}.

In terms of $\alpha_1$, the cold axion energy density is given
by Eq.~(\ref{3.41}).  We rewrite that equation here, assuming
$h \simeq 0.7$:
\beq
\Omega_a \simeq 0.15
\left({f_a \over 10^{12}~{\rm GeV}}\right)^{7 \over 6}\alpha_1^2~~~\ .
\label{case1}
\eeq
It has been remarked by many authors, starting with S.-Y. Pi \cite{Pi84},
that it is possible for $f_a$ to be much larger than $10^{12} {\rm GeV}$
because $\alpha_1$ may be accidentally small in our visible universe.
The requirement that $\Omega_a < \Omega_{\rm CDM} = 0.22$ implies
\beq
|{\alpha_1 \over \pi}| < 0.4
\left({10^{12}~{\rm GeV} \over f_a}\right)^{7 \over 12}~~~\ .
\label{alpha1}
\eeq
Since $- \pi < \alpha_1 < + \pi$ is the a-priori range of
$\alpha_1$ values and no particular value is preferred over
any other, $|{\alpha_1 \over \pi}|$ may be taken to be the
``probability" that the initial misalignment angle has
magnitude less than $|\alpha_1|$.  (Strictly speaking, the
word probability is not appropriate here since there is only
one universe in which $\alpha_1$ may be measured.)  If
$|{\alpha_1 \over \pi}| = 2 \cdot 10^{-3}$, for example,
$f_a$ may be as large as $10^{16}$ GeV, which is often
thought to be the ``grand unification scale".

The presence of isocurvature perturbations constrains the
small $\alpha_1$ scenario in two ways \cite{Turn91}.  First,
it makes it impossible to have $\alpha_1$ arbitrarily small.
Using Eq.~(\ref{axpow}), one can show that the fluctuations 
in the axion field cause the latter to perform a random walk
\cite{Lind90} characterized by the property
\begin{equation}
{1 \over V} \int_V d^3 x 
<(\delta a (\vec{x}, t) - \delta a (\vec{0}, t))^2> = 
4 \pi H_I^2 \ln(R k_{\rm max})
\label{randw}
\end{equation}
where the integral is over a sphere of volume 
$V = {4 \pi \over 3}R^3$ centered at $\vec{x} = 0$, and 
$k_{\rm max}$ is a cutoff on the wavevector spectrum.  
$a_1^2$ cannot be smaller than the RHS of Eq.~(\ref{randw}) 
with $R$ equal to the size of the present universe and 
$k_{\rm max}$ equal to the Hubble rate at QCD time, redshifted 
down to the present.  Since $\Omega_a < 0.22$, this implies 
a bound on $H_I$.  Translated to a bound on the scale 
of inflation $\Lambda_I$, defined by 
$H_I^2 = {8 \pi \over 3} G \Lambda_I^4$, it is
\beq
\Lambda_I < 5 \cdot 10^{14} {\rm GeV}
\left({f_a \over 10^{12}~{\rm GeV}}\right)^{5 \over 24}~~~\ .
\label{con1}
\eeq
Second, one must require axion isocurvature perturbations to be
consistent with CMBR observations.  Combining Eqs.~(\ref{isocon})
and (\ref{case1}), and setting $\Omega_{\rm CDM} = 0.22$,
${\delta \rho_m \over \rho_m} = 4.6~10^{-5}$, one obtains 
\beq
\Lambda_I < 10^{13} {\rm GeV}~~\Omega_a^{-{1 \over 4}}~
\left({f_a \over 10^{12}~{\rm GeV}}\right)^{5 \over 24}~~~\ .
\label{con2}
\eeq
Let us keep in mind that the bounds (\ref{con1}) and (\ref{con2}) 
pertain only if the reheat temperature $T_{\rm RH} < T_{\rm PQ}$.  
One may, for example, have $\Omega_a = 0.22$, $f_a \simeq 10^{12}$ GeV, and 
$\Lambda_I \simeq 10^{16}$ GeV, provided $T_{\rm RH} \gtwid 10^{12}$ GeV,
which is possible if reheating is sufficiently efficient.

\section{Acknowledgments}

It is a pleasure to thank Maria Beltr\'an, Juan Garcia-Bellido, 
Julien Lesgourgues, David Lyth, Michael Turner and Richard Woodard 
for enlightening discussions.  This work was supported in part by 
the U.S. Department of Energy under grant number DE-FG02-97ER41029.

%%%%%%%%%%%%%%%%%%%%%%%% referenc.tex %%%%%%%%%%%%%%%%%%%%%%%%%%%%%%
% sample references
% "physics"
%
% Use this file as a template for your own input.
%
%%%%%%%%%%%%%%%%%%%%%%%% Springer-Verlag %%%%%%%%%%%%%%%%%%%%%%%%%%

%
% BibTeX users please use
% \bibliographystyle{}
% \bibliography{}

\end{document}